\makeatletter
\def\ps@headings{%
\def\@oddhead{\mbox{}\scriptsize\rightmark \hfil \thepage}%
\def\@evenhead{\scriptsize\thepage \hfil \leftmark\mbox{}}%
\def\@oddfoot{}%
\def\@evenfoot{}}
\makeatother
\documentclass[journal,double-column]{IEEEtran}

\usepackage{latexsym,bm,amsmath,amssymb}
\usepackage[numbers]{natbib}
\usepackage{bm}
\usepackage{ifpdf}
\ifCLASSINFOpdf
   \usepackage[pdftex]{graphicx}
\else
   \usepackage[dvips]{graphicx}
\fi
\usepackage{subfigure}

\usepackage{bm}
\usepackage{url}

\begin{document}
%
\title{Who is Smarter? Intelligence Measure of Learning-based Cognitive Radios}



\author{\IEEEauthorblockN{Monireh Dabaghchian\IEEEauthorrefmark{1}, Amir Alipour-Fanid\IEEEauthorrefmark{1}, Songsong Liu\IEEEauthorrefmark{1}, Kai Zeng\IEEEauthorrefmark{1}, Xiaohua Li\IEEEauthorrefmark{2}, Yu Chen\IEEEauthorrefmark{2}}\\
\IEEEauthorblockA{\IEEEauthorrefmark{1}Volgenau School of Engineering,  George Mason University, Fairfax, VA 22030\\
\IEEEauthorrefmark{1}Email: \{mdabaghc, sliu23, aalipour, kzeng2\}@gmu.edu\\
\IEEEauthorrefmark{2}Electrical and Computer Engineering, Binghamton University, Binghamton, NY 13902\\
\IEEEauthorrefmark{2}Email: \{xli, ychen\}@binghamton.edu}
}

\maketitle

\begin{abstract}
Cognitive radio (CR) is considered as a key enabling technology for dynamic spectrum access to improve spectrum efficiency.
Although the CR concept was invented with the core idea of realizing ``cognition'', the research on measuring CR cognitive capabilities and intelligence is largely open.
Deriving the intelligence measure of CR not only can lead to the development of new CR technologies, but also makes it possible to better configure the networks by integrating CRs with different cognitive capabilities.
In this paper, for the first time, we propose a data-driven methodology to quantitatively measure the intelligence factors of the CR with learning capabilities.
The basic idea of our methodology is to run various tests on the CR in different spectrum environments under different settings and obtain various performance data on different metrics. Then we
apply factor analysis on the performance data to identify and quantize the intelligence factors and cognitive capabilities of the CR.
More specifically, we present a case study consisting of $144$ different types of CRs.
The CRs are different in terms of learning-based dynamic spectrum access strategies, number of sensors, sensing accuracy, processing speed, and algorithmic complexity.
Five intelligence factors are identified for the CRs through our data analysis. We show that these factors comply well with the nature of the tested CRs, which validates the proposed intelligence measure methodology.
\end{abstract}



%
\IEEEpeerreviewmaketitle

\section{Introduction}
 \label{intro}
In order to resolve the imminent spectrum shortage problem, sharing spectrum with legacy systems has attracted intensive research during the past decade.
Cognitive radio (CR), which has the capability to sense, learn, and adapt to the spectrum environment \cite{HLZH2011, MDAA2016, MDAA2017}, can significantly improve spectrum efficiency and guarantee the unharmful coexistence with the legacy systems \cite{AkyildizWL2006, QZBM2007, KSHK2010, MM2012, SDRF2008}.
Nevertheless, the complex and uncertain spectrum environment makes spectrum sharing extremely challenging.
The uncertainty may come from the radio propagation environment, the legacy system activity, or the complex behavior of the CR itself.

Just like human being, sophisticated cognitive capabilities are essential for the CR to cope with the uncertainty of spectrum environment.
The cognitive capabilities collectively define the intelligence of CR.
Although the CR concept was born with the core idea of realizing “cognition” \cite{ICAO2012}, the research on measuring CR cognitive capabilities or intelligence is largely open.

Being able to quantitatively measure the intelligence of CR can bring us a lot of benefits.

\begin{enumerate}

\item With the intelligence model and measuring methodology, we will gain deeper insight about the key factors that affect the intelligence of a CR which can be used to guide the development of new CRs with high intelligence.
\item A CR vendor may advertise and price their CR products based on CR intelligence as a metric.
A CR with higher intelligence tends to achieve better performance in practically uncertain spectrum environments, thus will be priced higher.
\item With the knowledge of the intelligence of individual CRs, a service provider or network manager can better configure their networks by integrating CRs with different intelligence levels in a more cost-efficient way.
For example, a CR with higher intelligence leading a set of CRs with lower intelligence may achieve a desirable performance with low network deployment cost.
\item Last but not the least, the investigation of CR intelligence will shed light on the intelligence measure of other smart systems, such as connected cars \cite{AAMD2017A, AAMD2017B}, unmanned aerial vehicles \cite{DM2015}, smart grid \cite{VGDS2011}, smart cities \cite{AZNB2014}, etc.
\end{enumerate}

This work is an extension of our previous work \cite{MDSL2016}, in which we proposed a data-driven methodology to derive the intelligence measure. We construct a CR intelligence model 
following human intelligence theory, specifically the widely accepted Cattell-Horn-Carroll (CHC) intelligence model \cite{KM2009chc}.
Based on this model, we develop psychometric techniques to measure the CR intelligence.
The basic idea of our methodology is to use simulations to test different CRs in various spectrum environments under different settings.
Based on the obtained performance data, we apply the factor analysis (FA) technique \cite{BMSP1952} to extract and measure the intelligence factors of CR.

More specifically, we present a case study consisting of 144 different types of CRs.
We provide each CR with different levels of capabilities including learning-based algorithms \cite{MDAA2017, Auer2002, Auer2003, Watkins1992} for dynamic spectrum access, number of sensors, sensing accuracy, processing speed, and algorithmic complexity.
With our methodology, five intelligence factors are identified for the CRs through our analysis, which are shown to comply with the nature of the tested algorithms. This validates our proposed methodology of measuring CR intelligence.

We summarize the contributions of this paper as follows:

\begin{itemize}

\item For the first time, we propose the idea of identifying the cognitive capabilities of CR and introduce an intelligence model for the CR.

\item We propose a methodology to extract the CR's intelligence factors and apply factor analysis as a theoretical framework for this purpose.

\item The proposed methodology is verified through a case study where we identify the intelligence factors of learning-based CRs under dynamic spectrum access scenarios and show these factors comply with the nature of the CRs.

\end{itemize}

The rest of the paper is organized as follows.
Section \ref{IQModel} proposes our intelligence model for CR.
Section \ref{IQMeasure} presents our methodology of deriving CR intelligence factors.
In section \ref{simulation}, we present a case study in which we measure the intelligence of learning based CRs under a dynamic spectrum access scenarios.
Section \ref{relWrk} discusses the related work and compares them with our approach. In particular, work on human intelligence measure and the difference between CR intelligence measure and human intelligence measure are highlighted.
Future work and open problems are discussed in Section \ref{Discussion}.
Section \ref{Conclsion} concludes the paper.

\section{Quantitative Intelligence Model of CR}
\label{IQModel}

\begin{figure}
						  \centering
							\includegraphics [scale = 0.4]{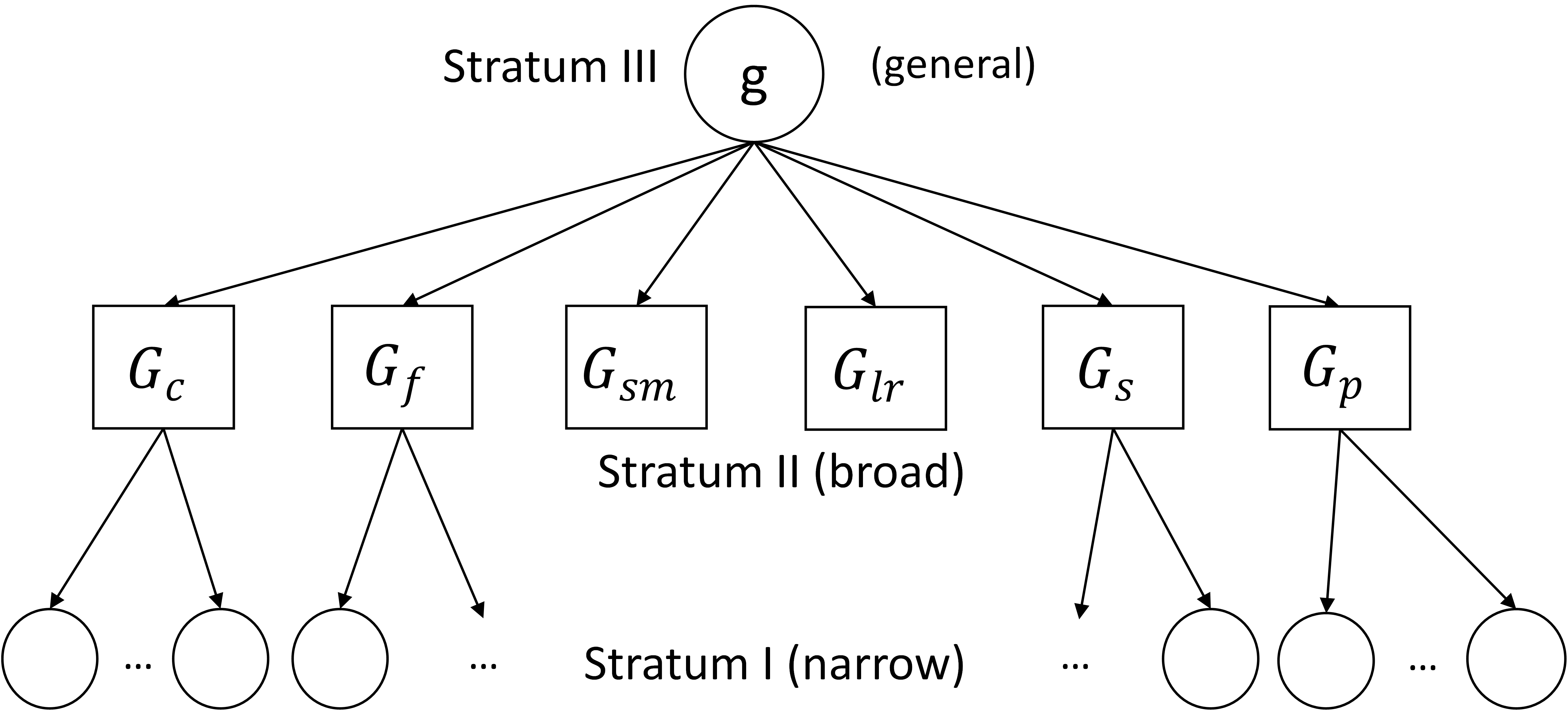}
						\caption{Intelligence model for cognitive radios.}
				   	\label{CRIntllgncMdlfig}
				\end{figure}
				
Motivated by the CHC model \cite{KM2009chc} that is widely used to describe human intelligence, we propose an intelligence model for the CR.
Our model is structured with three strata (or stages) as shown in Fig. \ref{CRIntllgncMdlfig}.
At the top stage lies the stratum III, which defines a unique general intelligence factor $g$.
CRs with higher values (or loadings) in the $g$ factor are more intelligent in general.
They tend to achieve better performance in various uncertain environments.

The stratum II represents a list of broad cognition capabilities contributing to intelligence, which are modeled as the following:
\begin{enumerate}
\item Comprehension-Knowledge ($G_{c}$): includes the breadth and depth of a CR's acquired knowledge and the ability to reason using previously learned experiences or procedures.

\item Fluid Reasoning ($G_{f}$): includes the broad ability to reason, form concepts, and perform dynamic spectrum access using unfamiliar information or novel procedures.

\item Short-Term Memory ($G_{sm}$): is the ability to apprehend and hold information in immediate awareness and then use it within a short period (e.g., a few seconds or the time the CR is on).

\item Long-Term Storage and Retrieval ($G_{lr}$): is the ability to store information and retrieve it later in the process of communication or dynamic spectrum access.

\item Spectrum Sensing ($G_{s}$): is the ability to sense the spectrum environment, e.g., sensing the availability of white space or presence of primary users.

\item Processing Speed ($G_{p}$): measures the information processing time, which includes delays resulted from channel sensing, accessing and switching, computing, reasoning, and information retrieval, etc.
The processing speed mainly refers to the delay or processing time required due to hardware limitations.

\item Algorithmic Processing Time ($G_{a}$) : is the time complexity of the algorithm employed within the cognitive radio.
It is also called algorithmic complexity. Different learning algorithms introduce different time complexity depending on the efficiency of the algorithms applied.
Note that algorithmic complexity is different from processing speed.

\end{enumerate}

Within each stratum II broad cognitive capability, we can further define stratum I, which is at the bottom, with more narrow and specific cognitive capabilities.
For example, fluid reasoning includes inductive reasoning, sequential reasoning, deductive reasoning, and speed of reasoning.
Spectrum sensing takes into account number of sensors and accuracy of sensing capability.
Processing speed considers the speed of processing on the received data, and the speed of switching among channels.
Algorithmic processing time consists of the speed of reasoning and decision making.

\section{Proposed Methodology to Measure the intelligence of CR}
\label{IQMeasure}

	      \begin{figure}
						  \centering
							\includegraphics [scale = 0.3]{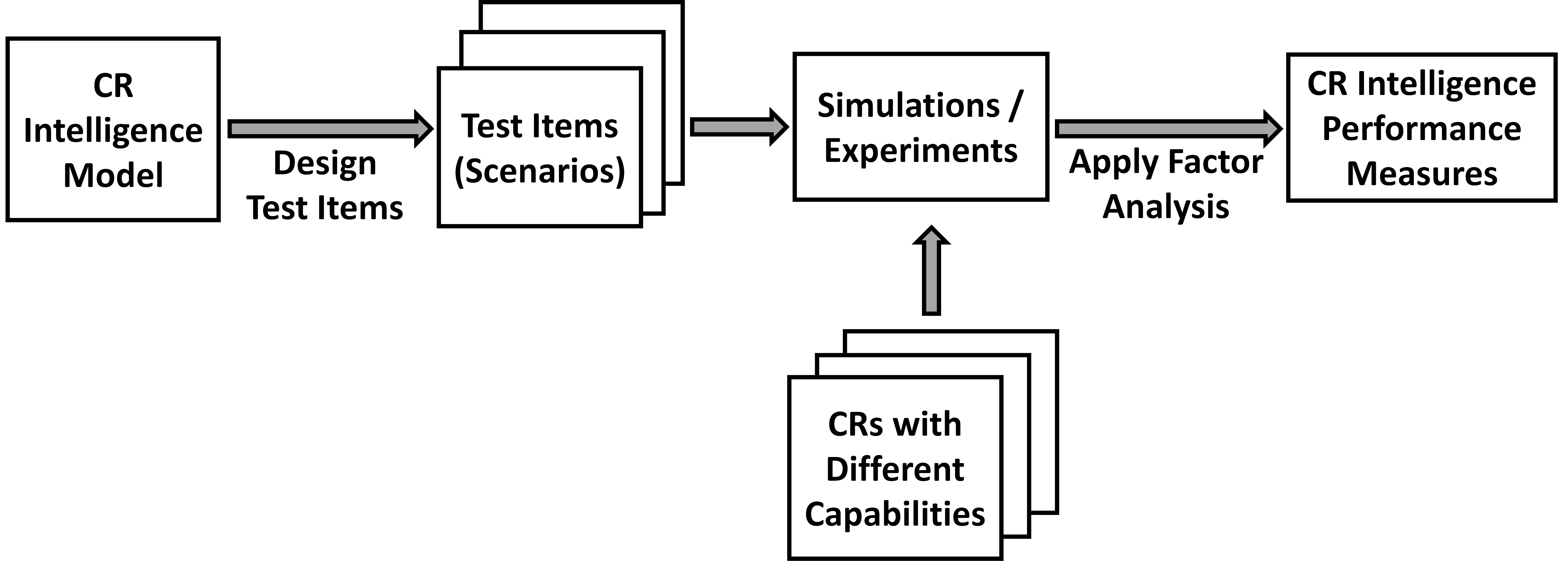}
						\caption{A data-driven methodology to measure the intelligence of CR.}
				   	\label{CRIntllgncMsure}
				\end{figure}
				
In this section, we propose a data-driven methodology to measure the intelligence of CRs.
The basic idea of this methodology is illustrated in Fig. \ref{CRIntllgncMsure}.

For a pool of $N$ different CRs called $CR_{1}$, $CR_{2}$, ..., $CR_{N}$, we design a set of $K$ test items to evaluate their performance.
CRs are different in terms of learning based spectrum access strategy, number of sensors, processing speed, computational complexity, etc. 
Various test environments arise from different primary user activity types or statistics, channel rates, frame delivery ratio, etc.
Through testing each CR in the testing scenarios, we obtain a vector of performance data $\bm{Y}_{k}(n)$ for each $CR_{n}$ ($1\leq n\leq N$) at each test scenario $k$ ($1\leq k\leq K$). 
The dimension of $\bm{Y}_{k}(n)$ equals to the number of performance metrics used. 
In our case study, $\bm{Y}_{k}(n)$ is an array of length three for each cognitive radio performing in a given test scenario since we measure three performance metrics for each CR.
Then we apply the FA method \cite{BMSP1952} on the measured data to derive the intelligence factors as latent factors. These factors are then matched to the broad cognitive capabilities described in Section \ref{IQModel} through analyzing the nature of the CR functions. 	

FA technique is applied on the data matrix $\bm{Y} = \{ \bm{Y}_k(n), \quad 1\leq n\leq N , 1\leq k\leq K \}$  , which identifies the latent factors as intelligence factors. The latent factors are then matched to the right cognitive capabilities by analyzing the functions of the CRs. 


There are two types of FA in the literature: exploratory FA and confirmatory FA \cite{BMSP1952,MSAF2009}.
Exploratory FA is used to identify the potential latent factors when both the number and the loading of the latent factors are unknown. 
Meanwhile, confirmatory FA is used when the number of latent factors are known. 
Then by applying the confirmatory FA we can decide whether the model and FA results match with each other or not.
It can also be used to test a theory on possible cognitive capabilities.
In other words, it determines whether or not the designed questions of the test measure the same factors that the questions were designed for.
In this paper, 
we use confirmatory FA to test our theory on the possible intelligence factors.

To describe the details about the intelligence model and the latent factors,
consider the performance of a test taker modeled as
\begin{equation}
y_{k}(n) = a_{k}g(n)+z_{k}(n),
\label{Equ201}
\end{equation}
where $y_k(n)$ is the measured performance of the cognitive radio $n$ on the testing scenario $k$, 
$g(n)$ is the general intelligence factor (see the stratum III of the intelligence model in Fig. \ref{CRIntllgncMdlfig}) of the cognitive radio $n$. The parameter $g(n)$ is called the ``common factor'', whose value determines how smart the CR $n$ is to achieve high performance value $y_k(n)$. 
The weighting coefficient $a_k$ denotes the loading, i.e., the importance, of the intelligence factor $g(n)$ on achieving high score $y_k(n)$ on the testing scenario $k$.
The value of $z_k(n)$ summarizes performance deviation from the simplified model $a_k g(n)$, which is unique to the specific performance measurement and is thus called the ``unique factor''.
Equation (\ref{Equ201}) also shows how cognitive capabilities or intelligence factors can be modeled by the common factor $g(n)$ \cite{BMSP1952}. 
Having all the measured data $y_k(n)$, we can use FA to determine whether the data fit the model ( Eq.~(\ref{Equ201}) and if so to estimate the loading $a_k$ and the intelligence factor $g(n)$. 

For more detailed cognitive capability analysis, we can consider the list of broad cognitive capabilities in stratum II. Let $x_i(n)$ denote the $i$th intelligence factor (or latent factor), where $1 \leq i \leq I$. The performance data vector $\bm{Y}_k(n)$ can be modeled as
\begin{equation}
\label{equ101}
\bm{Y}_{k}(n) = \bm{a}_{k,1}x_{1}(n) + \bm{a}_{k,2}x_{2}(n) + ... + \bm{a}_{k,I}x_{I}(n) + \bm{Z}_{k}(n),
\end{equation}
where $\bm{a}_{k,1}, \cdots, \bm{a}_{k,I}$ and $\bm{Z}_{k}(n)$ are the weights (loadings) and the unique factor, respectively.
Note that since it is possible to measure several metrics, the single value $y_k(n)$ in (\ref{Equ201}) is substituted by the vector performance measurement $\bm{Y}_{k}(n)$.
In this case, with all the measured data $\bm{Y}_k(n)$, we can verify the validity of the model (\ref{equ101}) and determine the weighting coefficients $\bm{a}_{k,i}$ as well as the latent factors $x_i(n)$. By analyzing the CR functioning, we can match the latent factors $x_i(n)$ with the CR stratum II cognitive capabilities listed Section \ref{IQModel}.

The FA technique \cite{BMSP1952} is applied to extract the group of latent factors $x_i(n)$ and then construct the CR intelligence model.
To apply the FA method, we rewrite Eq.~(\ref{equ101}) into the matrix form
\begin{equation} \label{eq1.1}
\bm{Y} = \bm{\Lambda}\bm{X} + \bm{\Psi},
\end{equation}
where $\bm{X}$ and $\bm{\Psi}$ are the matrices of common and the unique latent factors, respectively, and 
$\bm{\Lambda}$ is the matrix of weights $\bm{a}_{k,i}$. Specifically, 
\begin{align}
  \bm{Y}&=\left[\begin{array}{ccc} \bm{Y}_1(1) & \cdots & \bm{Y}_1(N) \\ \vdots & & \vdots \\ \bm{Y}_K(1) & \cdots & \bm{Y}_K(N) \end{array} \right], \nonumber \\
  \bm{\Lambda} &= \left[ \begin{array}{ccc} \bm{a}_1(1) & \cdots & \bm{a}_1(I) \\ \vdots & & \vdots \\
      \bm{a}_{K}(1) & \cdots & \bm{a}_{K}(I) \end{array} \right],
\end{align}
and the other matrices can be obtained similarly.

From Eq.~(\ref{eq1.1}), we can obtain the correlation matrix of the observation $\bm{Y}$ as
\begin{equation}
\label{equ104}
\bm{\Sigma} = \bm{E}\left(\bm{Y}\bm{Y}'\right)
              = \bm{\Lambda}\bm{\Phi}\bm{\Lambda'} + \bm{E}\left(\bm{\Psi}\bm{\Psi'}\right)
\end{equation}
where $\bm{\Phi} = E \left(\bm{X}\bm{X'} \right)$, and $E(\cdot)$ and $(\cdot)'$ denotes expectation and transposition, respectively.
The Eq. (\ref{equ104}) is derived based on the assumption that the common factor and unique factor are uncorrelated which yields $E\left(\bm{X}\bm{\Psi}'\right) = 0$. Similarly, based on the uncorrelation assumption, 
$\bm{E}\left(\bm{\Psi}\bm{\Psi'}\right)$ can be substituted by a diagonal positive definite matrix $\bm{\Gamma^{2}}$.
Therefore, Eq. (\ref{equ104}) can be rewritten as
\begin{equation}
\label{equ105}
\bm{\Sigma} = \bm{\Lambda}\bm{\Phi}\bm{\Lambda'} + \bm{\Gamma^{2}}.
\end{equation}

Without loss of generality, it is assumed that the latent factors $x_i(n)$ are orthogonal in the model. 
As a result $\bm{\Phi} = \bm{I}$. 
Then we subtract $\bm{\Gamma^{2}}$ from both sides of Eq.~(\ref{equ105}) to derive
\begin{equation} \label{equ51}
\bm{\Sigma} - \bm{\Gamma^{2}} = \bm{\Lambda}\bm{\Lambda'}.
\end{equation}
In this model, $\bm{\Sigma} - \bm{\Gamma^{2}}$ is called ``the reduced correlation matrix'' \cite{MSAF2009}.

The next step is to determine both $\bm{\Gamma^{2}}$ and $\bm{\Lambda}$.
Note that $\bm{\Gamma^2}$ is a diagonal matrix. If both $\bm{\Sigma}$ and $\bm{\Gamma^2}$ are known, then $\bm{\Lambda}$ can be estimated as $\bm{\Lambda} = \bm{A}\bm{D^{\frac{1}{2}}}$, where $\bm{A}$ is the eigenvector matrix and $\bm{D}$ is the diagonal eigenvalue matrix of the matrix $\bm{\Sigma} - \bm{\Gamma^{2}}$. 
On the other hand, if $\bm{\Lambda}$ has been estimated, then we can calculate $\bm{\Gamma^{2}}$ as
\begin{equation}
\bm{\Gamma^{2}} = \bm{\Sigma} - \bm{\Lambda}\bm{\Lambda'}.
\end{equation}

Therefore, with an initial estimate of $\bm{\Gamma}^2$, the Eq.~(\ref{equ51}) can be solved iteratively where each iteration involves the following three steps:

\begin{enumerate}

\item Find the eigenvector and eigenvalue matrices $\bm{A}$ and $\bm{D}$ of ``the reduced correlation matrix'': $\bm{\Sigma} - \bm{\Gamma^{2}} = \bm{A}\bm{D}\bm{A'}$;

\item Find $\bm{\Lambda} = \bm{A}\bm{D^{\frac{1}{2}}}$;

\item Find $\bm{\Gamma^{2}} = \bm{\Sigma} - \bm{\Lambda}\bm{\Lambda'}$.

\end{enumerate}

This procedure runs iteratively until the maximum difference of the last two round of $\bm{\Gamma^{2}}$ is less than certain small threshold \cite{MSAF2009}.

Let $\bm{S} = \bm{\Sigma} - \bm{D}$, then $\bm{\Sigma} - \bm{S}^{2}$ will generate the unrotated factors matrix.
Normally, we will pick up as latent factors those entries in $\bm{D}$  that are large enough, e.g., greater than 1.
In practice, we may simply use principal component analysis \cite{MSAF2009} to estimate $\bm{\Lambda}$, which just considers the latent factors influencing the performance and ignores the unique factors.

\section{Case Study: Intelligence Measure of CR with Learning Capabilities}
\label{simulation}

In this section, we present a case study consisting of different types of CRs.
By designing a set of testing environments, we apply our methodology presented in Section \ref{IQMeasure} to derive the latent factors and analyze them as intelligence factors as well as cognitive capabilities contributing to the CR intelligence.

        \begin{figure}
						  \centering
							\includegraphics [scale = 0.45]{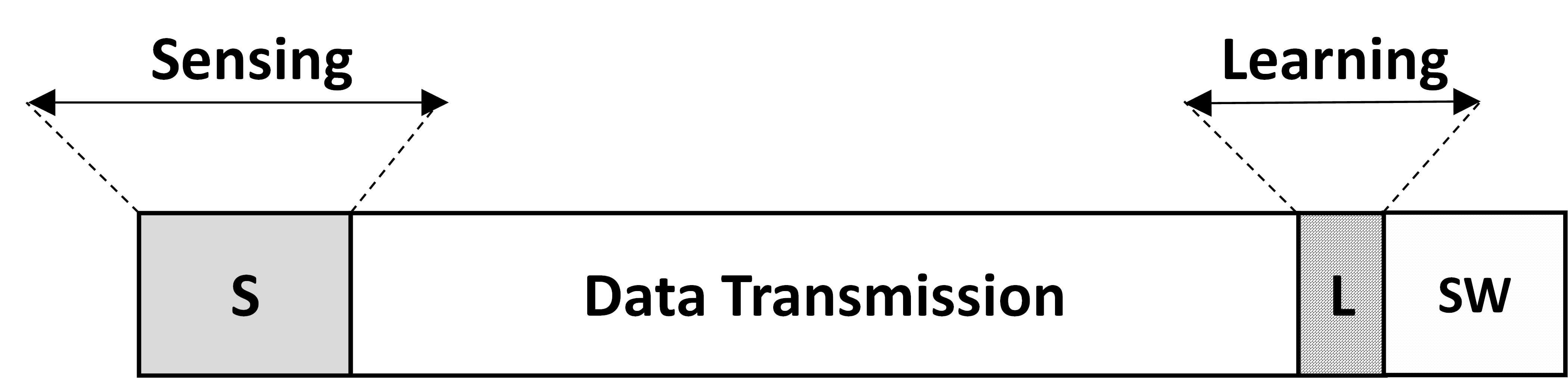}
						\caption{Time slot structure applied by the CR.}
				   	\label{timeslotfig}
				\end{figure}
				
\subsection{Settings}
\label{sttng}
We consider a single hop scenario where there is only one CR and one PU. Therefore, we can focus on each CR's performance without considering channel contention.
There are several channels in the network.
The PU can appear on some or all of the channels simultaneously.
We also assume a time slot based network.
Figure \ref{timeslotfig} shows the time slot structure used by the CR.

As shown in the figure, the first part of the time slot is assigned for channel sensing.
During this period, the CR senses the chosen channel and at the end of this period decides whether the channel is idle or not.
If the CR finds the channel idle, it begins data transmission.
Otherwise, it keeps silent to avoid interfering with the PU.

During the third part of the time slot, the CR learns from its observation.
No matter the channel was idle or busy, both of them provide useful information for the CR to learn and optimize its decisions in the future.
The last part is the switching period which indicates the amount of time that it takes the CR to switch from one channel to another one.
Switching period is dependent on the hardware limitations of each CR.

We have conducted extensive simulations with $144$ different types of CRs.
$10$ channels are considered in the network.
$18$ testing scenarios are designed, such that each CR performs on each of them one by one.
We run the simulations in MATLAB. For each CR performing in one single testing scenario we run the algorithm $10000$ times and get the average.
				
\subsection{Cognitive Radio Capabilities}

\begin{figure}
						  \centering
							\includegraphics [scale = 0.23]{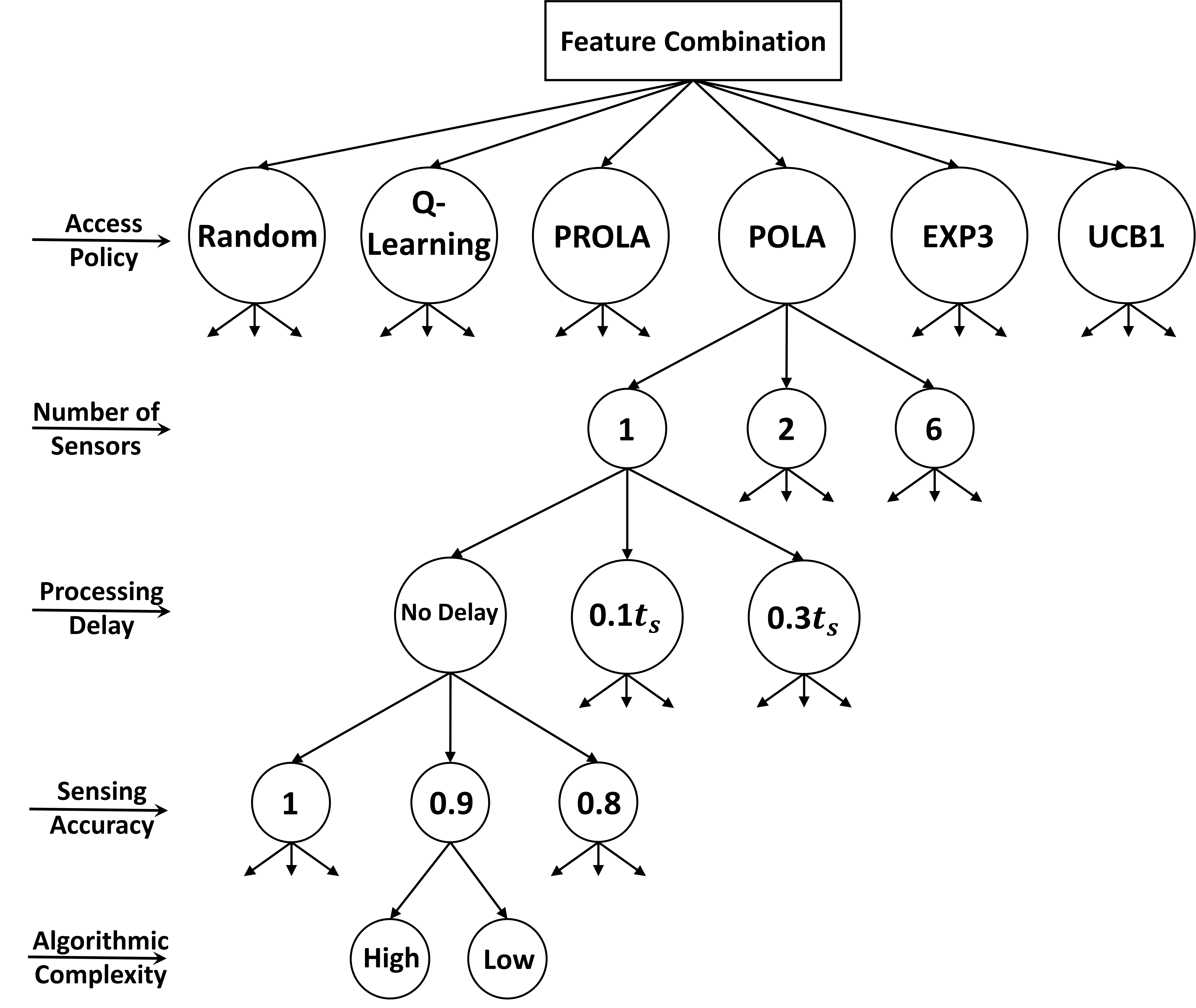}
						\caption{CRs consist of combinations of different features and parameters}
				   	\label{figCR}
\end{figure}	

Figure \ref{figCR} shows the capabilities of CRs considered in this case study in terms of their features and parameters. 
Combinations of all these features gives us 144 different types of CRs as explained in the following.
The CR features are described as follows.

\begin{itemize}

\item Channel access strategy (Access Policy) employed by the CR to learn and adapt to the environment.
It can be a learning-based method, deterministic or just a random strategy.
We consider five types of learning-based access strategies known as UCB1 \cite{Auer2002}, EXP3 \cite{Auer2003}, POLA\cite{MDAA2017}, PROLA\cite{MDAA2017}, and Q-Learning \cite{Watkins1992} and one random access strategy. Details of the strategies will be described in the sequel.

\item Number of sensors. Possessing more sensors, the CR observes more channels at each time slot. Then depending on the reasoning it employs, the CR may adapt better to the environment. This is probably equal to higher loads in cognitive capabilities. In this case study, we consider the number of sensors ($m$) to be either $1$, $2$, or $6$.

\item Sensing accuracy which indicates the detecting probability when the PU is present. There are several methods of channel sensing including energy detection and feature extraction \cite{SACT2011, WZRM2009, WCHK2012}. We consider three values of $1$, $0.9$, $0.8$ as the probability of the correct sensing. The values are relatively large because in practice, the CRs usually have high sensing accuracy.

\item Processing speed is another feature of a CR that occurs during sensing, learning, and switching parts of the time slot.
Learning delay occurs due to two reasons, the hardware limitations and due to algorithmic complexity of the learning algorithm.
We add up the delay due to hardware limitations that happen in different parts of a time slot as one single total delay. We assume this total delay to be either $0$, $0.1t_s$, or $0.3t_s$ in which $t_s$ indicates the time slot duration.

\item Algorithmic complexity. The delay occurred due to the time required by the computations in the algorithmic side is different than the delay due to hardware limitations.
It depends on the efficiency of the learning algorithm and for this reason it is called algorithmic complexity. 
This type of delay depends on how well the learning algorithm has been designed algorithmic-wise and it is inherent to the learning technique.
\end{itemize}

As to the six channel access strategies we employ in this work, the random access strategy does not utilize any learning-based algorithm. 
The other learning based algorithms mentioned are described below.

The UCB1 and EXP3 algorithms \cite{Auer2002, Auer2003} are slightly modified from their original version for the case with $m$ observations to address the more general case of observation of more than one channel. 
The modified UCB1 and EXP3 algorithms are described in Algorithms 1 and 2, respectively. 
Note that UCB1 is a deterministic access policy designed for well behaved environments, 
while EXP3 is designed for adversarial environments.

\noindent\rule{8.9cm}{0.5pt}\\
\textbf{Algorithm 1: UCB1 Algorithm with multiple observations}\\
\noindent\rule{8.9cm}{0.5pt}

\hangindent=1 em
\hangafter=1
\textbf{Initialization:} Play each machine once.
Per each play make $m$ observations including the played one.
The $m$ observations are made on the $m$ subsequent actions beginning from the action played.

\hangindent=1 em
\hangafter=1
\textbf{For each} $t = 2,...,T$: 
Play each machine that maximizes a given deterministic policy.
The decision criteria is based on the upper confidence bound concept from statistics.
Make $m$ observations on the $m$ subsequent channels beginning from the taken action.

\noindent\rule{8.9cm}{0.5pt}

\noindent\rule{8.9cm}{0.5pt}\\
\textbf{Algorithm 2: EXP3 Algorithm with multiple observations }\\
\noindent\rule{8.9cm}{0.5pt}

\hangindent=7.2 em
\hangafter=1
\textbf{Initialization:} Assign a uniform random distribution on action selection.

\textbf{For each} $t = 1,...,T $

\hangindent=2.4 em
\hangafter=1
1. Update the distribution on action selection based on the observations made so far plus adding some randomness.
Randomness is added to make sure the agent makes enough explorations.

\hangindent=2.4 em
\hangafter=1
2. Choose an action randomly based on the distribution defined above.

\hangindent=2.4 em
\hangafter=1
3. Observe the reward on $m$ subsequent channels beginning from the taken action.

\hangindent=2.4 em
\hangafter=1
4. Update the observation history on all the channels.
The observation history will be utilized in step one to optimize the channel selection distribution.

\noindent\rule{8.9cm}{0.5pt}

Algorithms 3 and 4 represent the POLA and the PROLA algorithms \cite{MDAA2017}. 
Both algorithms are designed for adversarial environments. 
PROLA as explained in Section \ref{relWrk} is similar to the EXP3 algorithm in the sense that at each time step, the agent is able to both gain reward and also to make an observation utilized in its learning process.
The difference between PROLA and EXP3 is that in EXP3, the agent observes the reward on the same action it takes and gains reward;
however, in PROLA, the agent makes observation on a channel other than the one it takes.

POLA is similar to the PROLA algorithm since both algorithms are designed to address the case when agent does not have the capability to observe the reward on the action it takes.
However, POLA has a major difference from the PROLA and EXP3 based on which at each time step, it can either take action or make observation.
This scenario, happens when the agent has limited capabilities and it cannot take action and switch to another channel for observation, during the periot of the same time step \cite{MDAA2017}.

\noindent\rule{8.9cm}{0.5pt}\\
\textbf{Algorithm 3: POLA Algorithm with multiple Observations}\\
\noindent\rule{8.9cm}{0.5pt}

\hangindent=7.2 em
\hangafter=1
\textbf{Initialization:} Assign uniform random distribution on the channels.

\textbf{For each} $t = 1,2,...,T $

\hangindent=2.4 em
\hangafter=1
1. With small probability $\epsilon$ decaying in time, choose an action uniformly at random to observe its reward.
Otherwise, take an action. 

\hangindent=2.2 em
\hangafter=1
2. If it is decised to make observation, choose $m$ channels to observe then update the channel selection probability based on the channel observation history.
Otherwise, choose a channel to access (take action) and accumulate the unobservable reward.

\noindent\rule{8.9cm}{0.5pt}

\noindent\rule{8.9cm}{0.5pt}\\
\textbf{Algorithm 4 : PROLA Algorithm with multiple Observations}\\
\noindent\rule{8.9cm}{0.5pt}

\hangindent=7 em
\hangafter=1
\textbf{Initialization:}
Assign random uniform distribution on channel selection.

\textbf{For each} $t = 1,2,...,T$ 

\hangindent=2.2 em
\hangafter=1
1. Assign a distribution on action selection based on the channel observation history.

\hangindent=2.2em
\hangafter=1
2. Choose a channel based on the above distribution to play and accumulate the unobservable reward. 

\hangindent=2.2em
\hangafter=1
3. Choose $m$ channels other than the played one uniformly at random to observe their reward during the same time slot.

\hangindent=2.2em
\hangafter=1
4. Update the channel observation history to optimize the distribution on channel selection policy.

\noindent\rule{8.9cm}{0.5pt} \\

The last learning algorithm we apply is Q-Learning algorithm \cite{ML1994} as described in Algorithm 5.
Q-Learning is similar to the UCB1 algorithm in the sense that they both are designed for well behaved environments.
More specifically, Q-learning algorithm is usually applied in the environments that follow a Markovian Chain.
One major difference between the Q-learning Algorithm and the UCB1 is that, Q-learning algorithm solves an optimization problem at each time step to optimize the action selection distribution.

In order to implement Q-Learning in MATLAB and to solve the optimization problems of this algorithm, CVX toolbox \cite{MGSB2008, MGSB2014} is used.
More specifically, CVX toolbox is designed to solve convex optimization problems in MATLAB.

Considering all the combinations of the features as shown in Fig.~(\ref{figCR}), $162$ different types of CRs are generated.
However, for random access strategy, no learning capability is utilized.
So the number of channels being observed makes no impact on the CR's performance.
By removing eighteen redundant combinations, $144$ CRs remain.
Different features and their assigned values are shown in Fig.~\ref{figCR}.

\noindent\rule{8.9cm}{0.5pt}\\
\textbf{Algorithm 5: Q-Learning with multiple observations }\\
\noindent\rule{8.9cm}{0.5pt}

\hangindent=7.2 em
\hangafter=1
\textbf{Initialization:}
Assign a random uniform distribution on channel selection.

\textbf{For each} $t = 1,2,...,T $

\hangindent=2.2 em
\hangafter=1
1. With an small probability choose an action uniformly at random to play. \\
Otherwise, choose an action with the distribution assigned based on the observation history.

\hangindent=2.2 em
\hangafter=1
2. Receive the reward on the action.
Make $m-1$ more observations on the subsequent channels other than the played one.

\hangindent=2.2 em
\hangafter=1
3. Use linear programming to optimize the action selection distribution.

\noindent\rule{8.9cm}{0.5pt}
			
\subsection{Testing Scenarios}

\begin{figure}
					\centering
							\includegraphics [scale = 0.26]{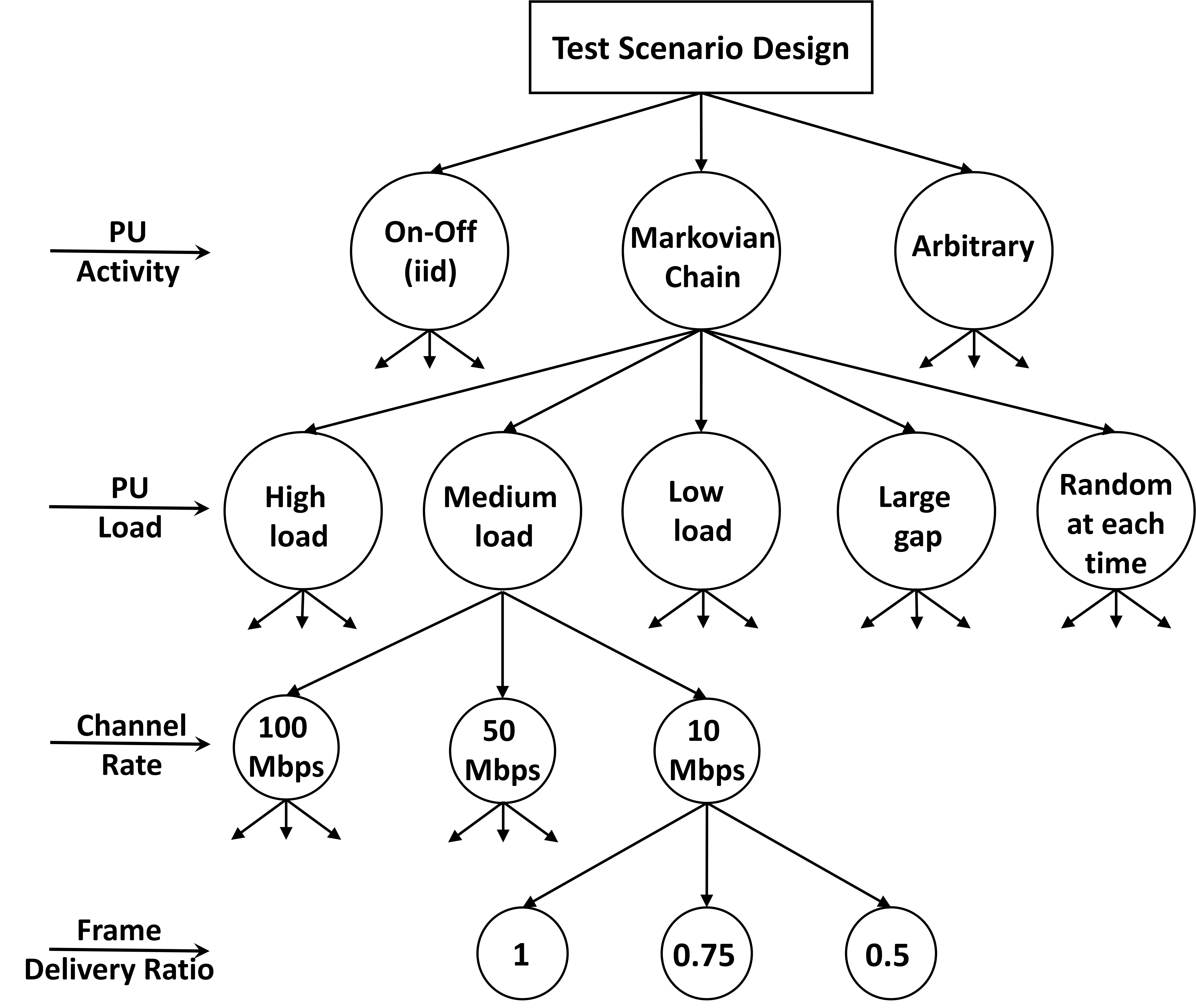}
						\caption{Designing Test Scenarios}
						\label{Envfig}
\end{figure}

We consider several parameters to design the testing scenarios:
\begin{itemize}

\item Type of PU Activity.
We consider three types of activities for the PU which consists of i.i.d. distribution, Markovian Chain, and arbitrary where no well defined distribution exists.

\item PU Load which indicates the probability of the PU to be active on each channel.
PU may have a high load on all the channels or may have a light load on only one channel and a heavy load on all other channels (large gap).
This testing scenario can discriminate among learning and nonlearning-based access strategies since by utilizing the observations and learning one can discriminate the good channel from low rewarding ones.
We have considered several combinations of PU activity on the channels.

\item Channel Rate. Three different values are considered as channel rates as shown in Fig.~\ref{Envfig}.
If we assume all other characteristics of the channels to be identical, a CR that learns the high rate channel may be considered as having high load in the corresponding cognitive capability.

\item Frame Delivery Ratio (FDR) which includes the impact of channel quality and noise on a given channel. Three possible values for FDR are considered in this case study.

\end{itemize}
Figure~\ref{Envfig} shows a summary of the parameters considered.
Combining these parameters, we create $18$ test scenarios.
Each CR needs to perform on each testing scenario so that its cognitive capabilities can be derived.

\subsection{Performance Metrics}


We measure the performance of the CRs based on three different metrics:

\begin{itemize}

\item Throughput which is stored as ${y}_{1k}(n)$ where $k$ and $n$ indicate the testing scenario and the CR indices, respectively.

\item Delay which indicates total delay occurred in the time slot and is stored as ${y}_{2k}(n)$. 

\item Violation ratio which represents the average number of times the CR interfered with the PU due to wrong sensing result called miss detection. It is assumed if the CR interferes with the PU, there will be a penalty for the CR and its data will be blocked, so there will be no throughput for the CR. Violation ratio is stored in ${y}_{3k}(n)$. 
\end{itemize}

The performance measure data vector $\bm{Y}_k(n)$ is equal to $\bm{Y}_k(n) = \left[{y}_{1k}(n) \;  {y}_{2k}(n) \; {y}_{3k}(n) \right]$ for $n = 1 , \ldots, 144$ and $k = 1 , \ldots, 18$.

\subsection{Simulation Results}
In this subsection we represent the simulation results, and analyze the intelligence factors as well as the cognitive capabilities of the CRs.
We divide our simulations into several phases.
In the first phase, we consider the UCB1, EXP3, and Random access based CRs.
Associated with each of UCB1 and EXP3 policies, there are twenty-seven CRs according to Fig.~\ref{figCR}.
There are nine CRs utilizing the random access strategy.

\begin{figure}
						  \centering
							\includegraphics [scale = 0.55]{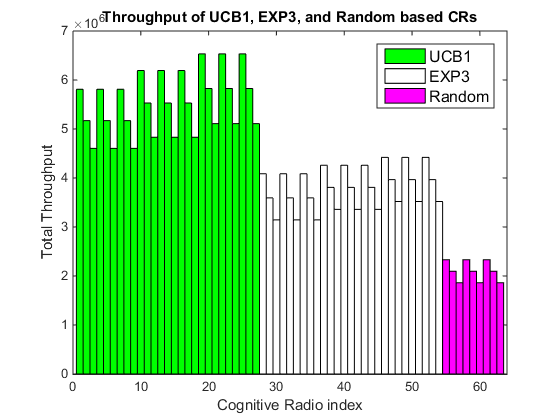}
						\caption{Total throughput of each CR achieved from all testing scenarios when the UCB1, EXP3, and Random access strategies are applied.}
				   	\label{thrghptfig}
				\end{figure}

Figure \ref{thrghptfig} shows the simulation result of the first metric, throughput.
This is the total throughput obtained by aggregating the throughput achieved from all the testing scenarios for each CR applying the mentioned access strategies.

From this figure, three clusters can be identified.
The first cluster (for cognitive radio index 1 to 27) represents CRs employing UCB1 learning-based access strategy.
The second cluster (for cognitive radio index 28 to 54) belongs to the CRs employing EXP3 learning-based access strategy.
The last cluster (for cognitive radio index 55 to 63) represents CRs utilizing random access strategies.

One observation is that, within each cluster, as the number of sensors increases, the overall throughput increases as well.
Next, the total throughput of CRs employing UCB1 is higher than those employing EXP3 since most of the testing scenarios designed are well behaved (stochastic) in which UCB1 performs better \cite{Auer2002,Auer2003}.
The third cluster illustrates those CRs employing random access methods.
Since random strategy never utilizes the previous observations, it achieves the lowest throughput among others.
The graphs also show that for each three consecutive CRs (i.e., three consecutive bars in the graph), the throughput is decreasing since the sensing accuracy is decreasing.

In the next step, we conduct data analysis via FA. From the simulations, three $63 \times 18$ matrices are generated for three metrics we measure. They all together create the data matrix $\bm{Y}$ with the dimension of $63\times 54$. FA is applied on this matrix using the software IBM SPSS \cite{wbstie}.

\begin{figure}
						  \centering
							\includegraphics [scale = 0.5]{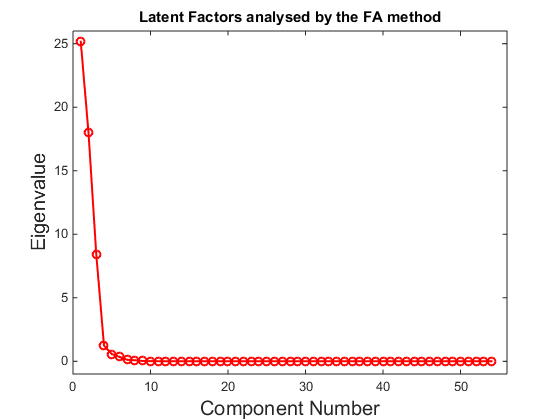}
						\caption{Latent factors identified for the UCB1, EXP3, and random-access based CRs based on the three metrics of throughput, delay, and violation ratio.}
				   	\label{FctrAnlsysfig}
\end{figure}

The analysis identifies four latent factors as shown in Fig. \ref{FctrAnlsysfig}.
Only four factors are distinguishable and the rest are negligible which are almost zero.
Due to limited space we skip the detailed output data corresponding to the FA results.
Even though the number of latent factors are identified, it is not yet clear which cognitive capabilities these factors correspond to.
We need to examine the data thoroughly and find out the corresponding cognitive capabilities by matching them to the CR functions. 	

By examining the data, the four latent factors (cognitive capabilities) are found as follows:
Spectrum sensing capability, processing speed capability, environment recognition capability, and environment adaptation capability.
The results are summarized in the first four rows of the Table \ref{fctrsTbl}.

As we study the results achieved by applying FA technique, the data of the first factor provides information on the violation ratio which is impacted by the sensing accuracy and the number of sensors.
As a result we conclude that the first latent factor corresponds to the spectrum sensing capability.
The second latent factor addresses the delay, which
is associated with the processing speed capability due to the hardware limitations of the CR.
The third factor is related to the learning capability, or specifically the environment recognition capability.
The forth factor shows a better performance for EXP3 and random access strategy than the UCB1 when the sensing accuracy decreases. The same thing happens when the environment is not well behaved.
This indicates that the EXP3 and random access strategy adapt better to non-well behaved environments.
The reason is because they utilize randomness in their access strategy which makes them more resilient to changes in the environment.
Deterministic based approaches assume a stable environment which makes them vulnerable to modifications in the environment.
As a result this latent factor addresses the environment adaptation capability.

Comparing to the intelligence model proposed in Section \ref{IQModel}, the processing speed capability matches the broad cognitive capability $G_p$, the spectrum sensing matches $G_s$, and the two others correspond to $G_c$ or $G_f$ as shown in Table \ref{fctrsTbl}.
In addition, all the CRs used in this case-study have high load on the $G_{sm}$ factor.

				\begin{figure}
						  \centering
							\includegraphics [scale = 0.5]{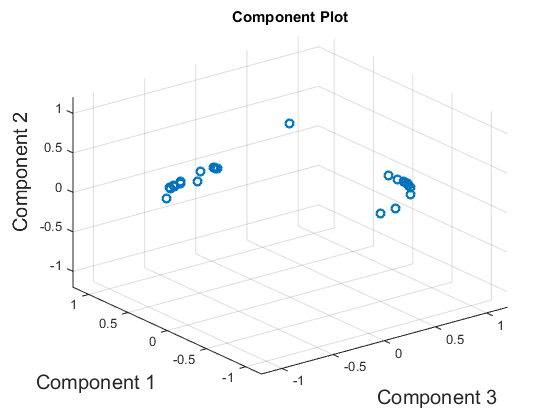}
						\caption{Component plot of the latent factors achieved by applying FA on all the three metrics.}
				   	\label{Compmtrixfig}
\end{figure}

\begin{figure}
						  \centering
							\includegraphics [scale = 0.5]{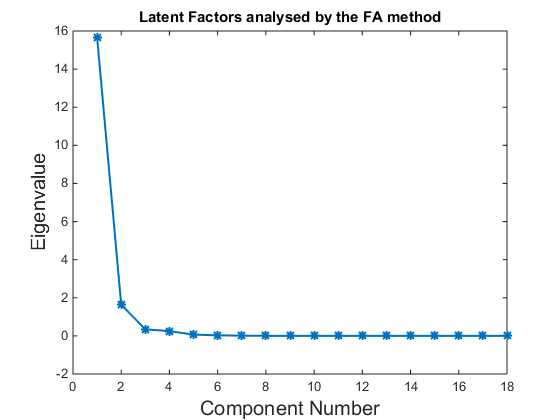}
						\caption{Latent factors identified for the UCB1, EXP3, and random-access based CRs based on only one metric, throughput.}
				   	\label{TwoFAfig}
				\end{figure}	
				
\begin{figure}
						  \centering
							\includegraphics [scale = 0.5]{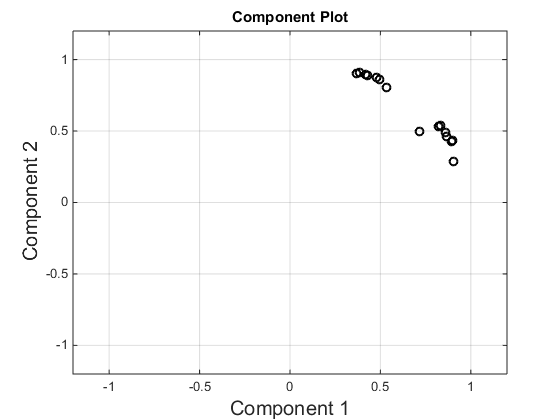}
						\caption{Component plot of the latent factors achieved by applying FA on the throughput metric.}
				   	\label{CompPlotfig}
				\end{figure}
				
\begin{table}[!t]
\renewcommand{\arraystretch}{1.2}
\caption{Latent factors identified that contribute to intelligence }
\label{fctrsTbl}
\centering
\begin{tabular}{l||l}
\hline  Factor I    &     Sensing Capability,     $G_s$         \\
\hline\hline   Factor II   &     Processing Speed Capability,    $G_p$         \\
\hline\hline   Factor III  &     Environment Recognition Capability,   $G_c$ or $G_f$          \\
\hline\hline   Factor IV   &     Environment Adaptation Capability,  $G_c$ or $G_f$      \\
\hline\hline   Factor V    &     Algorithmic Processing Time,       $G_a$ \\
\hline
\end{tabular}
\end{table}

Next, we plot the components obtained through the analysis. Component plot shows how the scenarios in the case study belong to each of the four latent factors. Since it is not possible to plot four dimensional figures, we plot the components for factors 1, 2 and 3 as shown in Fig. \ref{Compmtrixfig}. The whole data is divided into three clusters, each corresponding to one latent factor.

In order to get a deeper insight from the results, we also apply the FA technique to only one of the performance metrics called throughput.
In this case which is a limited case than the previous one, only two factors are identified as shown in Fig. \ref{TwoFAfig}.
One of them corresponds to the learning capability and the other one corresponds to the environment adaptation capability.
Figure \ref{CompPlotfig} shows the components of the analyzed data in which the whole data is divided into two clusters, each corresponding to one latent factor.

In the next phase of our simulation, we add the rest of the learning based CRs applying POLA, PROLA, and Q-Learning to the ones we considered earlier to make a comprehensive list of CRs with different capabilities.
Each of the $144$ CRs performs in the testing scenarios one by one.
Three performance metrics are measured.
This means that three matrices are generated, each with a dimension of $144\times 18$.
The combination of these matrices results in the data matrix $\bm{Y}$ with dimension $144 \times 54$.

\begin{figure}
						  \centering
							\includegraphics [scale = 0.55]{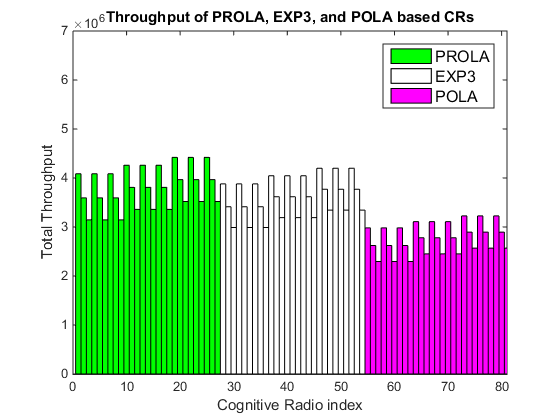}
						\caption{Total throughput of each CR achieved from all testing scenarios when the PROLA, EXP3, and POLA access strategies are applied.}
				   	\label{ThCmprsnEXP3POLAPROLAfig}
\end{figure}

As shown in Fig.~\ref{ThCmprsnEXP3POLAPROLAfig}, the performance of the PROLA is similar to the performance of the EXP3.
Algorithmic wise, the only difference between these two algorithms is that in EXP3, the agent observes the reward on the same action it takes;
while in the PROLA, the agent makes an observation on one other action different than the one it takes.
Our analysis shows that the cognitive capabilities of the PROLA is almost the same as the ones for EXP3.
All the three algorithms are designed for the non-stochastic environments.
As shown in the figure the POLA algorithm achieves a lower throughput compared to the two others. This is
because the POLA algorithm is not able to take action and make observation simultaneously at each time step.
Instead, it decides at each time step to do either of them.
This leads to a lower environment recognition capability and as a result POLA has a lower load in this cognitive capability compared to others.
In contrast, EXP3 and PROLA demonstrate almost equal loads with respect to this cognitive capability.
This indicates that non-stochastic based online learning algorithms do not necessarily demonstrate the same cognitive capabilities and should not be categorized into the same group.

\begin{figure}
						  \centering
							\includegraphics [scale = 0.55]{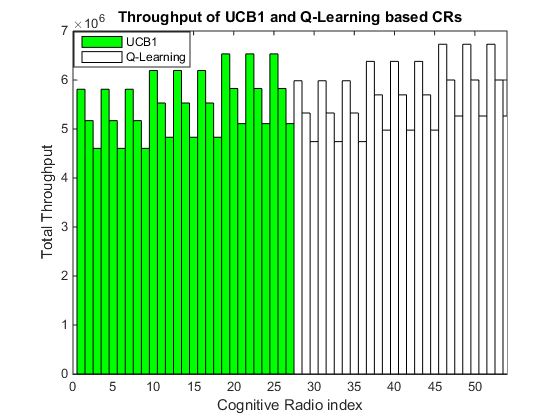}
						\caption{Total throughput of each CR achieved from all testing scenarios when the UCB1, Q-Learning access strategies are applied.}
				   	\label{UCBQLfig}
\end{figure}

\begin{figure}
						  \centering
							\includegraphics [scale = 0.5]{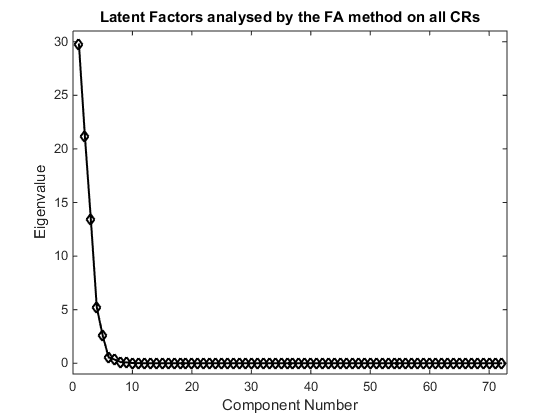}
						\caption{Latent factors identified considering all the CRs based on the three metrics of throughput, delay, and violation ratio.}
				   	\label{LatentfacotrALLfig}
\end{figure}		

Similarly, Fig.~\ref{UCBQLfig} shows the performance comparison of Q-learning and UCB1.
These two algorithms are both designed for stochastic environments.
As deterministic algorithms, they do not consider randomness in their policies.
Our results indicate that both algorithms show high loads in the cognitive capability of environment recognition.
However, their environment adaptability cognitive capability is low.
Q-Learning demonstrates low load in the cognitive capability of algorithmic processing.
This is because at each time slot, in order to update the action policy, the Q-learning algorithm solves an optimization problem.
In contrast, the UCB1 algorithm updates action policy at each time slot by a simple sum and multiplication operations.

Finally, we derive the latent factors as shown in Fig.~\ref{LatentfacotrALLfig}.
Five cognitive factors are identified with the fifth factor as the algorithmic processing time.
Table \ref{fctrsTbl} shows the whole list of factors identified in our case study.

\begin{figure*}[t]
\centering
\subfigure[Latent factors one, two, and three]{
\includegraphics[width=.31\textwidth]{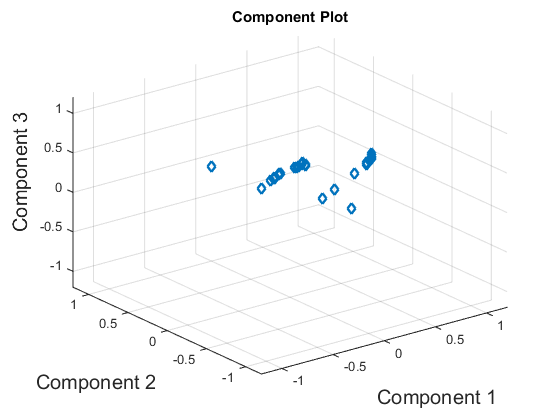}
}
\subfigure[Latent factors one, two, and four]{
\includegraphics[width=.31\textwidth]{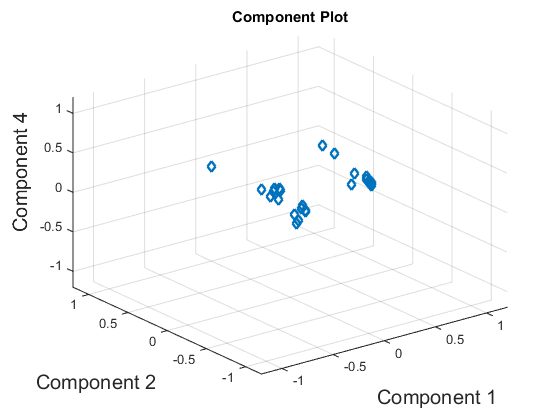}
}
\subfigure[Latent factors one, two, and five]{
\includegraphics[width=.31\textwidth]{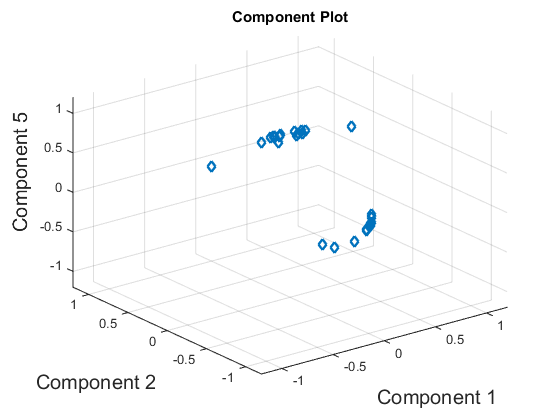}
}
\caption{Component plot of all five latent factors achieved by applying FA on all the three metrics.}
\label{COMPONENTPLOTALLfig}
\end{figure*}

We also show the component plot for the whole data set used in this case study in Fig.~\ref{COMPONENTPLOTALLfig}.
Since there are five latent factors, the component plot is five dimensional.
In order to represent the five dimensional data, we fix two of the latent factors, then plot three figures considering third, fourth, and fifth latent factors, respectively.

\section{Related Work}
\label{relWrk}

\subsection{Learning-based Algorithms}

\subsubsection{Multi Armed Bandits}

There is a rich literature about Multi Armed Bandits (MAB).
The MAB problems have many applications in cognitive radio networks with learning capabilities \cite{HLZH2011, MDAA2017, YGBK2010}.
In an MAB problem, an agent plays a machine repeatedly and obtains a reward when it takes a certain action at each time.
Any time when choosing an action the agent faces a dilemma of whether to take the best rewarding action known so far or to try other actions to find even better ones.
Trying to learn and optimize his actions, the agent needs to trade off between exploration and exploitation.
On one hand the agent needs to explore all the actions often enough to learn which is the most rewarding one and on the other hand he needs to exploit the believed best rewarding
action to minimize his overall regret.

MAB problems have been studied in different settings.
Stochastic MAB problems in which the rewards are generated i.i.d. on each arm are studied in \cite{Auer2002}.
The algorithm proposed is called UCB1 \cite{Auer2002}.
UCB1 is a deterministic strategy and the assumption is that the agent observes the reward on the action it takes.
The other algorithm with the same assumption of the reward observability on the taken action is called EXP3 \cite{Auer2003}.
EXP3 is designed for non-stochastic or adversarial environments when the adversary is oblivious.

There are also other types of MAB based on which the agent cannot observe the reward on the action it takes.
These MAB problems are usually called Multi Armed Bandits with side observations \cite{NANC2015}.
Two applications of MAB problems with side observations are addressed in \cite{MDAA2016, MDAA2017}.
PORLA algorithm is designed for the learning of an agent who takes action but instead of observing the reward on the action it took, it can observe the reward on any other action than the played one \cite{MDAA2017}.
Another algorithm called POLA is applied by an agent who not only cannot observe the reward on the action it plays but rather is not able to both take an action and then switch to another action to observe its reward at the same time step \cite{MDAA2017}.

\subsubsection{Reinforcement Learning}
Reinforcement learning (RL) is a branch of machine learning, designed for online learning.
Similar to MAB problems, the RL methods need to trade off between exploration and exploitation.
Q-Learning is a well studied topic and is categorized as a reinforcement learning technique that can be used to find the optimal action selection policy \cite{Watkins1992, ML1994}.
The environment is usually assumed to follow Markov Chain Process.

\subsection{Related Work on CR Intelligence}

Intelligence measure of CRs has not been well studied in the literature.
However, there are various studies on evaluating the performance of CRs.
A cognitive radio test methodology to test a CR system is presented in \cite{JTKH2015}.
The effect of cognitive engine on both SU and PU performance is measured and evaluated.
It is suggested that the cognition may be measured based on the SU's capability to improve its throughput and at the same time to decrease PU interference.
The authors call their method behavior-based testing.
In other words, their goal is to measure SU cognition based on the evaluation of both SU and PU performances instead of evaluating the SU cognition itself.
The testing scenarios are defined as narrow-band or wide-band environments.
The PU workloads and SU cognition considered in this work are limited and the authors suggest more research as a required step to justify the behavior-based cognition testing.
Statistical tools and the psychometrics are not utilized in contrast to our work that considers those methods.
This indicates that, our approach is completely different from this work.

The performance of cognitive radios is studied in \cite{AHFM2016} which considers four cognitive radio algorithms and intends to distinguish those that perform better than the others often enough.
They also study how sensitive different algorithms are to suboptimal parameters.
It is shown through simulations that, usually those algorithms that outperform others are highly sensitive to sub-optimal parameters.
While the others that show lower performance, represent a more steady performance and are more resistant to sub-optimality in the parameters.
The conclusion is that there is a trade-off between performance and consistency.
The difference of this work with ours is that their goal is to compare the performance of different
 learning based algorithms and to distinguish those that show consistent performance and have less dependency on the parameter values.
However, we derive the cognitive capabilities of CRs which is a totally different aspect of CR intelligence measurement.

\subsection{Cognitive Capabilities of Humans}

The cognitive capabilities and the intelligence model of human beings have been studied extensively in psychology \cite{RSSK2011}.
Human cognition capabilities include sensing, learning, memory, problem solving, etc. Intelligence
is defined as the ability to learn and perform cognitive tasks \cite{RSSK2011}.
Cattel-Horn-Carrot \cite{KM2009chc} is the most widely accepted model of human intelligence \cite{RSSK2011, SDRF2008}.

The practical measurement of mental abilities has been considered as a pivotal development in the behavioral sciences and the theories and techniques formed a field called “phychometrics”.
The first attempts of a mathematically more rigorous study of intelligence measure occurred in 1940s, with statistical techniques such as correlation and FA. Overall, FA is used in multiple areas including psychology and economics.

There have been some efforts trying to develop comprehensive benchmark frameworks to evaluate the cognitive radio network performance \cite{YZSM2009}, or to evaluate the performance of more general wireless networks \cite{NPSK,SRTT2011,GJTR2011}.
Since benchmarking wireless network is challenging, simulation has been adopted widely as a tool in the literature.
However, such benchmark studies are proposed not to test CR intelligence, but to evaluate CR performance.

It is helpful to identify the differences between human and CR intelligence capabilities.
One is that for human beings, the age of the test taker is an important factor that needs to be considered when designing the test questions, such as at the childhood stages in which the brain is still developing.
However, with respect to the CRs, a testing scenario can be tested by all types of CRs.

Another important difference is that a human being can get tired by the long duration of the test or may not be able to focus on the test day. This can make the test results unreliable. However, this is not a problem for CR and the test results can always be correct, unbiased and reliable.

\section{Discussion and Open Problems}
\label{Discussion}

\subsection{Cognitive Capabilities of Routing Algorithms}

Our current work is on the intelligence measure of CRs while they act in the MAC layer.
Intelligence measure of CRs in the routing layer is an interesting future research direction.
There are some preliminary work done on the learning-based routing methods \cite{AVMS2017}, where
the authors try to answer the question of whether machine learning including deep reinforcement learning can replace the traditional network protocol design.
It is shown that data driven based routing methods that extract information from the traffic history achieve better performance.
For any learning based routing algorithm designed for cognitive radio networks, we can measure their intelligence and cognitive capabilities, similarly. 
This leads to designing better routing algorithms and better network configurations to maximize network throughput while minimizing costs.

\subsection{Item Response Theory and IQ Measure}

After extracting intelligence factors and identifying cognitive capabilities of CRs, the next step would be to combine these capabilities and assign a quantitative value to it called Intelligence Quotient (IQ).
This is in fact the unique general intelligence factor g in Stratum III shown in Fig.~\ref{CRIntllgncMdlfig}.
In order to do so, one needs to first make sure that the test scenarios are comprehensive and standardized.
In other words, the testing items shown in Fig.~\ref{CRIntllgncMsure} should include all types of test scenarios from easy to hard ones.
Item Response Theory (IRT) \cite{SESR2000} which is a design, analysis, and scoring paradigm for tests, is the tool that needs to be used to quantify the easy and difficult test scenarios.
Using IRT to design the optimal test scenarios and to develop the IQ measurement methods is another interesting future research direction.

\subsection{Configuring the Network with Combination of CRs with Different Intelligence}

As explained in the introduction, cognitive radio networks can be configured by integrating CRs with different intelligence and cognitive capabilities.
This may lead to the optimal use of resources and would also be more cost efficient.
More comprehensive research is needed in order to quantitatively measure the performance of such networks and to rigorously show how one or a few number of CRs with higher intelligence
can lead and network with other CRs with lower intelligence.

\section{Conclusion}
\label{Conclsion}

In this paper, for the first time, we have proposed the idea of deriving the intelligence measure and analyzing the cognitive capabilities of the CR.
An intelligence model is proposed for the CR, and a data-driven methodology is proposed which applies FA techniques to identify CR intelligence factors and cognitive capabilities.
A case study is presented in which through extensive simulations, five latent factors are identified for the CR that comply well with the nature of the tested CRs.

Our ongoing effort is focused on measuring the intelligence quotient (IQ) for each CR. IQ can be considered as the general intelligence factor that indicates how well a CR performs in uncertain environments. We will also expand our methods to measure CR intelligence in multi-user and multi-hop networks.


\section*{Acknowledgment}

M. Dabaghchian, A. Alipour-Fanid, S. Liu, and K. Zeng are partially supported by the NSF under grant No. CNS-1502584, CNS-1464487, and CNS-1619073.
X. Li and Y. Chen are supported by NSF via grant CNS-1443885.




\bibliographystyle{IEEEtran}
%



\bibliography{Ref}

\end{document}